\documentclass[sigconf]{acmart}
\AtBeginDocument{%
  \providecommand\BibTeX{{%
    \normalfont B\kern-0.5em{\scshape i\kern-0.25em b}\kern-0.8em\TeX}}}

\copyrightyear{2023}
\acmYear{2023}
\setcopyright{acmlicensed}\acmConference[KDD '23]{Proceedings of the 29th ACM SIGKDD Conference on Knowledge Discovery and Data Mining}{August 6--10, 2023}{Long Beach, CA, USA}
\acmBooktitle{Proceedings of the 29th ACM SIGKDD Conference on Knowledge Discovery and Data Mining (KDD '23), August 6--10, 2023, Long Beach, CA, USA}
\acmPrice{15.00}
\acmDOI{10.1145/3580305.3599422}
\acmISBN{979-8-4007-0103-0/23/08}

\usepackage{booktabs}
\usepackage{multirow}
\usepackage{algorithm}
\usepackage{algorithmic}
\usepackage{enumitem}
\usepackage{xcolor}
\usepackage{subfigure}
\usepackage{diagbox}

\newcommand{\eg}{\emph{e.g.}}
\newcommand{\ie}{\emph{i.e.}}

\usepackage{pifont}

\begin{document}

\title{MAP: A Model-agnostic Pretraining Framework for Click-through Rate Prediction}

\author{Jianghao Lin}
\email{chiangel@sjtu.edu.cn}
\affiliation{
  \institution{Shanghai Jiao Tong University}
  \city{Shanghai}
  \country{China}
}
\author{Yanru Qu}
\authornote{Weinan Zhang and Yanru Qu are co-corresponding authors.}
\email{kevinqu16@gmail.com}
\affiliation{
  \institution{Shanghai Jiao Tong University}
  \city{Shanghai}
  \country{China}
}
\author{Wei Guo}
\email{guowei67@huawei.com}
\affiliation{
  \institution{Huawei Noah's Ark Lab}
  \city{Shenzhen}
  \country{China}
}
\author{Xinyi Dai}
\email{daixinyi@sjtu.edu.cn}
\affiliation{
  \institution{Shanghai Jiao Tong University}
  \city{Shanghai}
  \country{China}
}
\author{Ruiming Tang}
\email{tangruiming@huawei.com}
\affiliation{
  \institution{Huawei Noah's Ark Lab}
  \city{Shenzhen}
  \country{China}
}
\author{Yong Yu}
\email{yyu@sjtu.edu.cn}
\affiliation{
  \institution{Shanghai Jiao Tong University}
  \city{Shanghai}
  \country{China}
}
\author{Weinan Zhang}
\authornotemark[1]
\email{wnzhang@sjtu.edu.cn}
\affiliation{
  \institution{Shanghai Jiao Tong University}
  \city{Shanghai}
  \country{China}
}

\renewcommand{\shortauthors}{Jianghao Lin and Yanru Qu, et al.}

\begin{abstract}
With the widespread application of personalized online services, click-through rate (CTR) prediction has received more and more attention and research.
The most prominent features of CTR prediction are its multi-field categorical data format, and vast and daily-growing data volume.
The large capacity of neural models helps digest such massive amounts of data under the supervised learning paradigm, yet they fail to utilize the substantial data to its full potential, since the 1-bit click signal is not sufficient to guide the model to learn capable representations of features and instances.
The self-supervised learning paradigm provides a more promising pretrain-finetune solution to better exploit the large amount of user click logs, and learn more generalized and effective representations.
However, self-supervised learning for CTR prediction is still an open question, since current works on this line are only preliminary and rudimentary.
To this end, we propose a \textbf{M}odel-\textbf{a}gnostic \textbf{P}retraining (MAP) framework that applies feature corruption and recovery on multi-field categorical data, 
and more specifically, we derive two practical algorithms: masked feature prediction (MFP) and replaced feature detection (RFD).
MFP digs into feature interactions within each instance through masking and predicting a small portion of input features, and introduces noise contrastive estimation (NCE) to handle large feature spaces. 
RFD further turns MFP into a binary classification mode through replacing and detecting changes in input features, making it even simpler and more effective for CTR pretraining. 
Our extensive experiments on two real-world large-scale datasets (\ie, Avazu, Criteo) demonstrate the advantages of these two methods on several strong backbones (\eg, DCNv2, DeepFM), and achieve new state-of-the-art performance in terms of both effectiveness and efficiency for CTR prediction.

\end{abstract}



\maketitle

\section{Introduction}

\begin{figure}[t]
    \centering
    \includegraphics[width=0.48\textwidth]{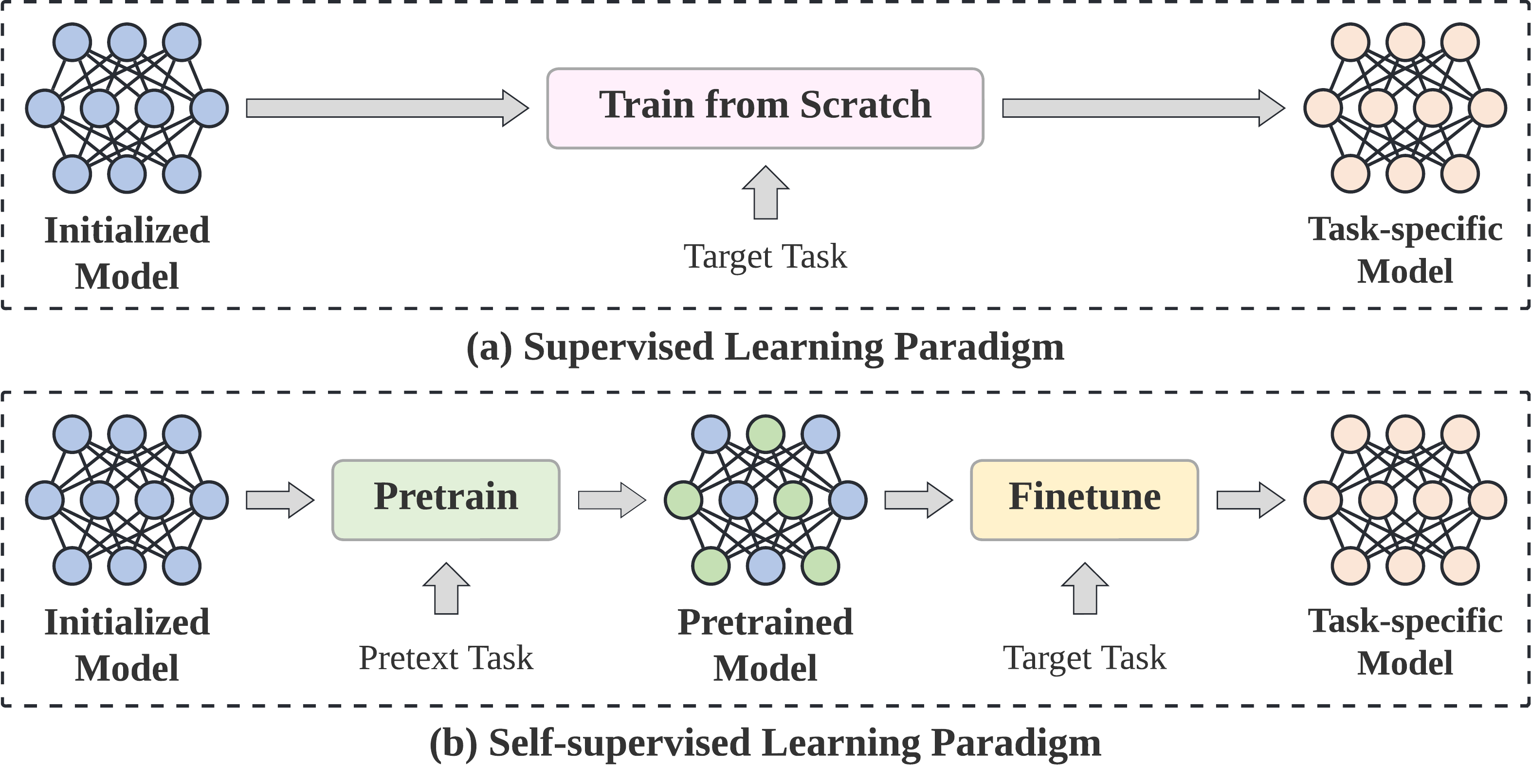}
    \vspace{-20pt}
    \caption{The illustration of (a) supervised learning paradigm, and (b) self-supervised learning paradigm.
    The supervised learning paradigm directly trains a randomly initialized model from scratch without pretraining. The self-supervised learning paradigm contains two stages, where we first pretrain the model based on the pretext task and then finetune it for the downstream task.
    }
    \vspace{-10pt}
    \label{fig:SL vs SS}
\end{figure}

Click-through rate (CTR) prediction aims to estimate the probability of a user's click~\cite{wang2022enhancing,huang2022neural,xi2023towards} given a specific context, and plays a fundamental role in various personalized online services, including recommender systems~\cite{xi2023bird}, display advertising~\cite{qin2020user}, web search~\cite{lin2021graph,dai2021adversarial,fu2023f}, etc.
Traditional CTR models (\eg, logistic regression~\cite{richardson2007predicting} and FM-based models ~\cite{he2017neural,rendle2012factorization}) can only capture low-order feature interactions, which might lead to relatively inferior performance in real-world applications.
With the rise of deep learning techniques and the massive amount of user behavior data collected online, many delicate neural CTR models have been proposed to model higher-order feature interactions with different operators (\eg, product~\cite{DCNv2,DeepFM,PNN,PIN,FiBiNET,xDeepFM}, convolution~\cite{FGCNN,CCPM,FiGNN}, and attention~\cite{AutoInt,InterHAt,AFM}).
These works generally follow a supervised learning paradigm shown in Figure~\ref{fig:SL vs SS}(a), where a model is randomly initialized and trained from scratch based on the supervised signals (click or not). Nevertheless, the 1-bit click signal is not sufficient enough for the model to learn capable representations of features and instances, resulting in suboptimal performance. 

Self-supervised learning provides a more powerful training paradigm to learn more generalized and effective representations of data samples, and proves to be effectual in Natural Language Processing (NLP)~\cite{devlin2018bert,clark2020electra} and Computer Vision (CV)~\cite{MoCo} domains. 
As shown in Figure~\ref{fig:SL vs SS}(b), they usually adopt a pretrain-finetune scheme, where we first pretrain an encoder based on pretext tasks (\ie, pretraining tasks), and then finetune a model initialized by the pretrained encoder for downstream tasks based on specific training data with supervised signals.
According to the pretext tasks, self-supervised learning can be mainly classified into two categories: (1) contrastive methods and (2) generative methods~\cite{liu2021self}.
Contrastive methods~\cite{MoCo,chen2020simple,oord2018representation,chen2020improved,grill2020bootstrap,bahri2021scarf} aim to learn generalized representations from different views or distortions of the same input.
Generative methods~\cite{devlin2018bert,he2022masked,brown2020language,yang2019xlnet} reconstruct the original input sample from the corrupted one.

Self-supervised learning has flourished in the NLP domain to pretrain transformers for unlabeled sentences, making it a perfect mimicry target for sequential recommendations. Various methods are proposed to treat user behavior sequences as sentences, and adopt language models~\cite{sun2019bert4rec} or sentence augmentations~\cite{liu2021contrastive,xie2020contrastive} for better user or item embeddings.
Moreover, many pretraining methods are designed for different types of data formats (\eg, graph data~\cite{wang2021pre,hao2021pre,meng2021graph}, or multi-modal data~\cite{liu2022multi,liu2021pre}) to further enrich the recommendation family.
However, these sequential/graph/multi-modal based pretraining methods are essentially incompatible for the CTR data format, \ie, multi-field categorical data format:
\begin{align}
x_i= \underset{Item=Jeans\,\,\,\,}{\underbrace{\left( 0,...,1,0 \right) }}\underset{Color=Blue\,\,}{\underbrace{\left( 1,...,0,0 \right) }}\mathbf{...}\underset{Gender=Male}{\underbrace{\left( 1,0 \right). }}
\label{eq:multi-field categorical data}
\end{align}

In this paper, we focus on self-supervised pretraining over multi-field categorical data for CTR prediction. 
There exist preliminary works~\cite{bahri2021scarf,wang2020masked} that explore the pretraining methods for CTR data. 
MF4UIP~\cite{wang2020masked} leverages the BERT framework~\cite{devlin2018bert} to predict the masked features for user intent prediction.
However, it is non-scalable when the feature space grows, and thus suffers from severe inefficiency problem for industrial applications with million-level feature spaces.
SCARF~\cite{bahri2021scarf} is a contrastive method that adopts SimCLR framework~\cite{chen2020simple} and InfoNCE loss~\cite{oord2018representation} to learn robust representations for each data sample. 
Its contrastive property requires to calculate representations from different views of the same instance, which doubles the throughput time and memory usage, leading to low efficiency.
Moreover, contrastive based methods only provide coarse-grained instance-level supervisions from sample pairs, and therefore might get trapped in representation spaces' early degeneration problem~\cite{liu2021self}, where the model overfits the pretext task too early and loses the ability to generalize. 

To this end, we propose a \textbf{M}odel-\textbf{a}gnostic \textbf{P}retraining (MAP) framework that applies feature corruption and recovery towards multi-field categorical data for CTR pretraining.
Specifically, we derive two algorithms based on different strategies of corruption and recovery: masked feature prediction (MFP) and replaced feature detection (RFD).
MFP requires the model to recover masked features according to corrupted samples, and adopts noise contrastive estimation (NCE) to reduce the computational overhead caused by the large feature space (million level) in CTR data.
Moreover, RFD turns MFP into a binary classification mode and requires the model to detect whether each feature in the corrupted sample is replaced or not.
Compared with MFP that only utilizes a subset of fields and predict over the entire feature space, RFD is simpler yet more effective and more efficient, which provides fine-grained and more diverse field-wise self-supervised signals.
RFD can achieve better CTR performance with fewer parameters, higher throughput rates, and fewer pretraining epochs compared to other pretraining methods.
Derived from the MAP framework, MFP and RFD are compatible with any neural CTR models and can promote performance without altering the model structure or inference cost. 

Main contributions of this paper are concluded as follows:
\begin{itemize}[leftmargin=10pt]
    \item We propose a Model-agnostic Pretraining (MAP) framework that applies feature corruption and recovery on multi-field categorical data. Different pretraining algorithms could be derived by customizing the strategies of corruption and recovery.
    \item We derive a masked feature prediction (MFP) pretraining algorithm from MAP, where the model predicts the original features that are replaced by <$\text{MASK}$> tokens.
    We also adopt noise contrastive estimation (NCE) to reduce the computational overhead.
    \item We derive a replaced feature detection (RFD) pretraining algorithm from MAP, where the model is required to detect whether the feature of each field is replaced or not. 
    RFD is simpler yet more effective and more efficient, and it can achieve better CTR performance with fewer computational resources.
    \item Extensive experiments on two real-world large-scale datasets validate the advantages of MFP and RFD on several strong backbones, and achieve new state-of-the-art performance in terms of both effectiveness and efficiency for
CTR prediction.
\end{itemize}
\section{Preliminaries}
\label{sec:preliminary}

Without loss of generality, the basic form of CTR prediction casts a binary classification problem over multi-field categorical data.
Each instance for CTR prediction contains $F$ fields with each field taking one single value from multiple categories, and can be represented by $\left\{ x_i, y_i \right\}$. $x_i$ is a sparse one-hot vector as shown in Eq.~\ref{eq:multi-field categorical data}, and $y_i\in{\{1,0\}}$ is the true label (click or not). 
For simplicity, we build a global feature map of size $M$, and assign a unique feature index for each category, and thus we can represent each sample as $x_i=[x_{i,1},\dots,x_{i,F}]$, where $x_{i,f}\;(f=1,\dots,F)$ is the index of corresponding feature.

\begin{figure*}[t]
    \centering
    \includegraphics[width=0.99\textwidth]{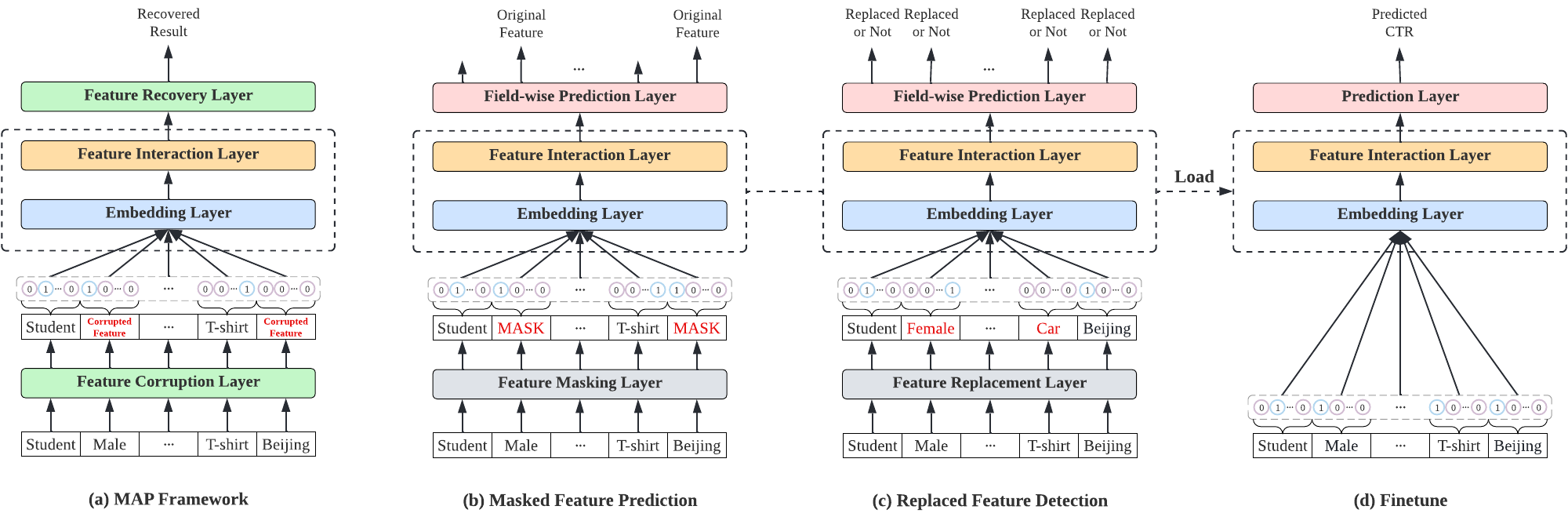}
    \caption{
    The illustration of (a) MAP framework, (b) masked feature prediction (MFP), (c) replaced feature detection (RFD), and (d) finetune. 
    MFP and RFD are derived from the MAP framework by customizing the design of feature corruption and recovery layers.
    In the finetuning stage, we maintain the same model structure, and load the parameters from the pretrained model to initialize the embedding layer and feature interaction layer.
    }
    \label{fig:framework}
\end{figure*}

CTR models aim to estimate the click probability $P(y_i=1|x_i)$ for each sample. 
According to~\cite{wang2022enhancing,zhang2021deep}, the structure of most recent CTR models can be abstracted as three layers: (1) embedding layer, (2) feature interaction layer, and (3) prediction layer.

\textbf{Embedding layer} transforms the sparse binary input $x_i$ into dense low-dimensional embedding vectors $\mathbf{E}=[v_1;v_2;\dots;v_F] \in \mathbb{R}^{F\times d}$, where $d$ is the embedding size, and each feature is represented as a fixed-length vector $v_f \in \mathbb{R}^{d}$.

\textbf{Feature interaction layer}, as the main functional module of CTR models, is designed to capture the second- or higher-order feature interactions with various operations (\eg, product, attention). This layer produces a compact representation $q_i$ based on the dense embedding vectors $\mathbf{E}$ for the sample $x_i$.

\textbf{Prediction layer} estimates the click probability $\hat{y}_i=P(y_i=1|x_i)$ based on the representation $q_i$ generated by the feature interaction layer. It is usually a linear layer or an MLP module followed by a sigmoid function:
\begin{equation}
\sigma(x)=\frac{1}{1+e^{-x}},
\end{equation}

After the prediction layer, the CTR model is trained in an end-to-end manner with the binary cross-entropy loss:
\begin{equation}
\mathcal{L} =-\frac{1}{N}\sum_{i=1}^N \left[y_i\log\hat{y}_i + (1-y_i)\log(1-\hat{y}_i)\right],
\end{equation}
where $N$ is the number of training samples.

\section{Methodology}
\label{sec:method}

In this section, we introduce our proposed Model-agnostic Pretraining (MAP) framework, and derive two pretraining algorithms based on different strategies of feature corruption and recovery. The illustration of the framework and algorithms is shown in Figure~\ref{fig:framework}.

\subsection{MAP Framework Overview}

We adopt the common pretrain-finetune scheme in self-supervised learning for NLP~\cite{devlin2018bert,clark2020electra} and CV~\cite{MoCo}, where we first pretrain a CTR model for a pretext task, and then finetune the pretrained model with click signals.

We propose a Model-agnostic Pretraining (MAP) framework for the pretraining stage. 
The pretext task for the model is to recover the original information (\eg, original features, corrupted field index) from the corrupted samples.
It is worth noting that MAP is compatible with any neural CTR models, since we only corrupt the input sample (\ie, feature corruption layer) and alter the prediction head (\ie, feature recovery layer) for the recovery target.
Finally, by customizing the design of feature corruption and recovery layers, we derive two specific pretraining algorithms as follows: 

\begin{itemize}[leftmargin=10pt]
    \item \textbf{Masked feature prediction (MFP)} requires the model to recover the original features from the corrupted sample which contains multiple <$\text{MASK}$> tokens.
    \item \textbf{Replaced feature detection (RFD)} tells the model to perform field-wise detection about whether the feature of each field is replaced or not.
\end{itemize}

Hereinafter, we omit the detailed structure of feature interaction layer for  certain CTR models, since MFP and RFD are both model-agnostic pretraining algorithms.

\subsection{Masked Feature Prediction}

In the masked feature prediction (MFP) pretraining stage, we first corrupt the original input sample $x_i$ with a feature masking layer, where we randomly replace a certain proportion of the features with <$\text{MASK}$> tokens. 
Then, we feed the corrupted sample $x^c_i$ through the embedding layer and feature interaction layer to get the compact representation $q_i^{c}$. 
Finally, the vector $q_i^{c}$ is inputted to a field-wise prediction layer to predict the original feature for each <$\text{MASK}$> token. 
To ensure efficiency and practicability, we introduce noise contrastive estimation (NCE) to allow the model to predict among a large feature space (\eg, millions of candidate features).

\subsubsection{Feature Masking Layer}

For an input sample with $F$ features (\ie, $x_i=[x_{i,1},\dots,x_{i,F}]$), we randomly replace a part of the features with <$\text{MASK}$> tokens, resulting in a corrupted sample $x_i^c$.
The proportion of features to be masked is a hyperparameter denoted as corrupt ratio $\gamma$.
We represent the set of indices of masked fields as $\mathcal{I}$.
The <$\text{MASK}$> token is also regarded as a special feature in the embedding table, and it is shared among all the feature fields.
That is, we do not maintain field-specific mask tokens, in order to avoid introducing prior knowledge about the masked fields.
A harder pretraining task with less prior knowledge can force the model to learn more generalized feature representations, which benefits the downstream task~\cite{clark2020electra,liu2019roberta}.

\subsubsection{Field-wise Prediction Layer}

After the embedding layer and feature interaction layer, we obtain the representation $q_i^c$ for the corrupted sample $x_i^c$. 
For each masked feature $x_{i,f}$ of the $f$-th field, we maintain an independent multi-layer perceptron (MLP) network $g_f$ followed by a softmax function to compute the predictive probability $p_{i,f}\in \mathbb{R}^{M}$ over the candidate features:
\begin{equation}
    z_{i,f}=g_f(q_i^c),\;z_{i,f}\in\mathbb{R}^{M},
\end{equation}
\begin{equation}
    p_{i,f,j}=\frac{\exp({z_{i,f,j}})}{\sum_{k=1}^{M} \exp(z_{i,f,k})},\;j=1,\dots,M.
    \label{eq:MFP softmax}
\end{equation}

We expand the predictive space (\ie, candidate features) for every masked field $f$ from the field-specific feature space to the global feature space, in order to increase the difficulty of pretext task and thus benefit the downstream CTR prediction task~\cite{clark2020electra,liu2019roberta}.
That is, the model has to select the original feature $x_{i,f}$ out of the whole feature space, which usually contains millions of features in recommender systems.
Finally, we view MFP pretraining as a multi-class classification problem and employ the multi-class cross-entropy loss for optimization:
\begin{equation}
    \mathcal{L}_i^{MFP}=\frac{1}{|\mathcal{I}|}\sum_{f\in\mathcal{I}} \operatorname{CrossEntropy}(p_{i,f},x_{i,f})
\end{equation}

\subsubsection{Noise Contrastive Estimation}

The MFP method introduced above is still impractical and extremely expensive, since in Eq.~\ref{eq:MFP softmax} we have to calculate the softmax function over the large global feature space. 
Such a million-level multi-class classification problem leads to a tremendous amount of memory usage and unacceptable pretraining time cost for real-world applications.
To this end, we adopt noise contrastive estimation (NCE)~\cite{gutmann2010noise,mikolov2013efficient,mnih2012fast} to reduce the softmax overhead.

NCE converts the multi-class classification problem into a binary classification task, where the model tries to distinguish the positive feature (\ie, the masked feature $x_{i,f}$) from noise features. 
Specifically, for the $f$-th masked field, we sample $K$ noise features from $M$ candidate features according to their frequency distribution in the training set. Then, we employ the binary cross-entropy loss:
\begin{equation}
    \mathcal{L}_i^{NCE}=-\frac{1}{|\mathcal{I}|}\left[\sum_{f\in\mathcal{I}} (\log\sigma(z_{i,f,t}) + \sum_{k=1}^K \log(1-\sigma(z_{i,f,k})))\right],
    \vspace{2pt}
\end{equation}
where $z_{i,f}$ is the output of $f$-th MLP predictor, $t$ is the feature index of the positive feature, and $\sigma$ is the sigmoid function.
In this way, we reduce the complexity of loss calculation from $O(M)$ to $O(Km)$, where $Km\ll M$, and $m$ is the number of masked fields.
In our experiment, $K=25$ is enough to achieve a good CTR performance for the large global feature space (\eg, $M=4\,\text{millions}$).


\subsection{Replaced Feature Detection}

As shown in Figure~\ref{fig:framework}(c), we further propose the replaced feature detection (RFD) algorithm to provide fine-grained and more diverse pretraining signals from all the feature fields, instead of a subset of fields in MFP (\ie, the masked fields).

In the RFD pretraining stage, we first corrupt the original input sample $x_i$ by a feature replacement layer, where we randomly replace a certain proportion of the features with other features.
Then, after obtaining the compact representation $q_i^c$ from the embedding and feature interaction layers, we employ a field-wise prediction layer to detect whether the feature in each field is replaced or not.

\subsubsection{Feature Replacement Layer}

For an input sample with $F$ features (\ie, $x_i=[x_{i,1},\dots,x_{i,F}]$), we randomly replace a part of the features, and denote the set of indices of replaced fields as $\mathcal{I}$. 
The proportion of features to be replaced is a hyperparameter represented as corrupt ratio $\gamma$.
Next, we replace each of these selected features by a random sampling from the empirical marginal distribution (\ie, sample from the field-specific feature space by the feature frequency distribution) of the corresponding field $\hat{\mathcal{F}}_f$ in the training set, resulting in the corrupted sample $x_i^c$. 

\subsubsection{Field-wise Prediction Layer}

Similar to MFP, we obtain the representation $q_i^c$ for the corrupted sample $x_i^c$ through the embedding layer and feature interaction layers.
Then, we feed $q_i^c$ to an MLP predictor followed by an element-wise sigmoid function, resulting in an $F$-length predictive vector $p_i$:
\begin{equation}
\begin{aligned}
p_i=\sigma(\operatorname{MLP}(q_i^c)),\;p_i\in\mathbb{R}^{F},
\end{aligned}
\end{equation} 
where $p_{i,f}\;(f=1,\dots,F)$ denotes the probability that the feature in the $f$-th field is replaced in the feature replacement layer. Finally, we employ the binary cross-entropy loss for RFD pretraining:
\begin{equation}
    \mathcal{L}_i^{RFD}=\frac{1}{F} \sum_{f=1}^F \operatorname{BinaryCrossEntropy}(p_{i,f},r_{i,f}),
    \vspace{2pt}
\end{equation}
where $r_{i,f}\in\{1,0\}\;(f=1,\dots,F)$ is the label indicating whether the feature in the $f$-th field of sample $x_i$ is replaced or not. 

Apparently, RFD serves as a simpler pretraining algorithm compared with MFP which requires NCE to reduce the computational overhead. 
While MFP only utilizes the masked fields (the corrupt ratio $\gamma$ is usually 10\% - 30\%) as self-supervised signals, RFD involves all the feature fields to provide more sufficient and more diverse signal guidance.
We will evaluate the effectiveness and efficiency of RFD later in Section~\ref{sec:exp effectiveness} and Section~\ref{sec:exp efficiency}, respectively.



\subsection{Complexity Analysis}
\label{sec:complexity analysis}

We analyze the time complexity of our proposed MFP and RFD, as well as two baseline pretraining algorithms (\ie, MF4UIP and SCARF).
We only analyze the component above the feature interaction layer (\ie, prediction layer and loss calculation) due to their model-agnostic property.

Suppose the batch size is $B$, and we adopt a linear layer $f:\mathbb{R}^{l}\rightarrow\mathbb{R}^n$ to transform the compact representation $q_i$ (or $q_i^c$) to the final predictive place.
The time complexity of MF4UIP is $O(Bln+BM)$, where the main overhead of MF4UIP is the softmax computation over the million-level feature space of size $M$.
By adopting NCE, the time complexity of MFP reduces to $O(Bln+BKm)$, where $Km\ll M$, and $m$ is the number of masked features.
By introducing the binary classification mode, the time complexity of RFD further reduces to $O(Bln+BF)$, where $F$ is the number of feature fields.
Besides, as a contrastive algorithm with InfoNCE loss, SCARF has a 
quadratic complexity over the batch size: $O(2Bln+4B^2n)$.

In summary, MF4UIP and SCARF are non-scalable in terms of feature space $M$ and batch size $B$, respectively. MFP and RFD can achieve lower complexity and is scalable for industrial applications with million-level feature space and large batch size.
\section{Experiment}

In this section, we conduct extensive experiments to answer the following research questions:
\begin{itemize}
    \item[\textbf{RQ1}] Can MFP and RFD improve the predictive performance for various base CTR models?
    \item[\textbf{RQ2}] How do MFP and RFD perform compared with existing CTR pretraining methods?
    \item[\textbf{RQ3}] Does RFD improve the pretraining efficiency compared with other pretraining methods?
    \item[\textbf{RQ4}] What are the influences of different pretraining configurations for MFP and RFD?
\end{itemize}

\subsection{Experiment Setup}
\subsubsection{Datasets}
We conduct extensive experiments on two large-scale CTR prediction benchmarks, \ie, Avazu and Criteo datasets.
Both datasets are divided into training, validation, and test sets with proportion 8:1:1. The basic statistics of these two datasets are summarized in Table~\ref{tab:dataset}.
Note that the training set for the pretraining stage and finetuning stage are the same, in order to make full use of the large-scale datasets.
We describe the preprocessing for the two datasets as follows:
\begin{itemize}[leftmargin=10pt]
    \item \textbf{Avazu} originally contains 23 fields with categorical features. We remove the \emph{id} field that has a unique value for each data sample, and transform the \emph{timestamp} field into four new fields: \emph{weekday}, \emph{day\_of\_month}, \emph{hour\_of\_day}, and \emph{is\_weekend}, resulting in 25 fields. We remove the features that appear less than 2 times and replace them with a dummy feature <$\text{Unknown}$>.
    \item \textbf{Criteo} includes 26 anonymous categorical fields and 13 numerical fields. We discretize numerical features and transform them into categorical features by log transformation\footnote{\url{https://www.csie.ntu.edu.tw/~r01922136/kaggle-2014-criteo.pdf}}. We remove the features that appear less than 10 times and replace them with a dummy feature <$\text{Unknown}$>.
\end{itemize}
\begin{table}[h]
    \centering
    \caption{The dataset statistics}
    \vspace{-10pt}
    \label{tab:dataset}
	\resizebox{0.48\textwidth}{!}{
	\renewcommand\arraystretch{1.1}
    \begin{tabular}{c|c c c c c}
    \toprule
    Dataset & \#Training & \#Validation & \#Test & \#Fields & \#Features \\
    \midrule
    Avazu & 32,343,172 & 4,042,897 & 4,042,898 & 25 & 4,428,327 \\
    Criteo & 36,672,493 & 4,584,062 & 4,584,062 & 39 & 1,086,794 \\
    \bottomrule
    \end{tabular}
    }
\end{table}

\subsubsection{Evaluation Metrics}
To evaluate the performance of CTR prediction methods, we adopt AUC (Area under the ROC curve) and Log Loss (binary cross-entropy loss) as the evaluation metrics. 
Slightly higher AUC or lower Log Loss (\eg, 0.001) can be regarded as significant improvement in CTR prediction~\cite{xDeepFM,DCNv2,wang2022enhancing}

\begin{table*}[t]
	\centering
	\caption{
	The AUC and Log Loss (LL) performance of different base CTR models under different training schemes. We give the relative performance improvement of each pretrain-finetune scheme over the ``scratch'' scheme.
	The best result for each base model is given in bold, while the second-best value is underlined. 
	The symbol * indicates statistically significant improvement of our proposed MFP and RFD schemes over the two baseline schemes with $p<0.001$. 
	}
    \vspace{-10pt}
	\label{tab:overall performance}
	\resizebox{\textwidth}{!}{
	\renewcommand\arraystretch{1.0}
	\begin{tabular}{c|c|c|c c c c|c c c c}
		\toprule
		\cline{1-11}
		\multicolumn{1}{c|}{\multirow{2}{*}{FI Operator}} & \multicolumn{1}{c|}{\multirow{2}{*}{Base Model}} &  \multicolumn{1}{c|}{\multirow{2}{*}{Scheme}} &  \multicolumn{4}{c|}{Avazu} & \multicolumn{4}{c}{Criteo}  \\
		\cline{4-11}
		\multicolumn{1}{c|}{} & \multicolumn{1}{c|}{} & \multicolumn{1}{c|}{} & AUC & $\Delta$AUC$\;\uparrow$ & LL & $\Delta$LL$\;\downarrow$ & AUC & $\Delta$AUC$\;\uparrow$ & LL & $\Delta$LL$\;\downarrow$ \\
		
		\cline{1-11}
		\multicolumn{1}{c|}{\multirow{5}{*}{MLP}} & \multicolumn{1}{c|}{\multirow{5}{*}{DNN}} & Scratch & 0.7920 & - & 0.3740 & - & 0.8105 & - & 0.4413 & - \\
		\multicolumn{1}{c|}{} & \multicolumn{1}{c|}{} & MF4UIP & 0.7991 & 0.90\% & 0.3683 & 0.0057 & 0.8135 & 0.37\% & 0.4386 & 0.0027 \\
		\multicolumn{1}{c|}{} & \multicolumn{1}{c|}{} & SCARF & 0.7989 & 0.87\% & 0.3684 & 0.0056 & 0.8122 & 0.21\% & 0.4399 & 0.0014 \\
		\multicolumn{1}{c|}{} & \multicolumn{1}{c|}{} & MFP (Ours) & $\;\underline{0.8006}^*$ & 1.09\% & $\;\underline{0.3675}^*$ & 0.0065 & $\;\underline{0.8145}^*$ & 0.49\% & $\;\underline{0.4373}^*$ & 0.0040 \\
		\multicolumn{1}{c|}{} & \multicolumn{1}{c|}{} & RFD (Ours) & $\;\textbf{0.8016}^*$ & 1.22\% & $\;\textbf{0.3666}^*$ & 0.0074 & $\;\textbf{0.8152}^*$ & 0.58\% & $\;\textbf{0.4367}^*$ & 0.0046 \\
		
		\cline{1-11}
		\multicolumn{1}{c|}{\multirow{10}{*}{Attention}} & \multicolumn{1}{c|}{\multirow{5}{*}{AutoInt}} & Scratch & 0.7895 & - & 0.3751 & - & 0.8068 & - & 0.4452 & - \\
		\multicolumn{1}{c|}{} & \multicolumn{1}{c|}{} & MF4UIP & 0.7919 & 0.30\% & 0.3729 & 0.0022 & 0.8089 & 0.26\% & 0.4429 & 0.0023 \\
		\multicolumn{1}{c|}{} & \multicolumn{1}{c|}{} & SCARF & 0.7951 & 0.71\% & 0.3708 & 0.0043 & 0.8084 & 0.20\% & 0.4435 & 0.0017 \\
		\multicolumn{1}{c|}{} & \multicolumn{1}{c|}{} & MFP (Ours) & $\underline{0.7952}$ & 0.72\% & $\underline{0.3707}$ & 0.0044 & $\;\underline{0.8100}^*$ & 0.40\% & $\;\underline{0.4420}^*$ & 0.0032 \\
		\multicolumn{1}{c|}{} & \multicolumn{1}{c|}{} & RFD (Ours) & $\;\textbf{0.7978}^*$ & 1.05\% & $\;\textbf{0.3693}^*$ & 0.0058 & $\;\textbf{0.8104}^*$ & 0.45\% & $\;\textbf{0.4416}^*$ & 0.0036 \\
		\cline{2-11}
		\multicolumn{1}{c|}{} & \multicolumn{1}{c|}{\multirow{5}{*}{Transformer}} & Scratch & 0.7922 & - & 0.3751 & - & 0.8071 & - & 0.4446 & - \\
		\multicolumn{1}{c|}{} & \multicolumn{1}{c|}{} & MF4UIP & 0.7945 & 0.29\% & 0.3713 & 0.0038 & 0.8090 & 0.24\% & 0.4427 & 0.0019 \\
		\multicolumn{1}{c|}{} & \multicolumn{1}{c|}{} & SCARF & 0.7958 & 0.45\% & 0.3705 & 0.0046 & 0.8096 & 0.31\% & 0.4422 & 0.0024 \\
		\multicolumn{1}{c|}{} & \multicolumn{1}{c|}{} & MFP (Ours) & $\;\underline{0.7968}^*$ & 0.58\% & $\;\underline{0.3700}^*$ & 0.0051 & $\;\underline{0.8112}^*$ & 0.51\% & $\;\underline{0.4406}^*$ & 0.0040 \\
		\multicolumn{1}{c|}{} & \multicolumn{1}{c|}{} & RFD (Ours) & $\;\textbf{0.8003}^*$ & 1.02\% & $\;\textbf{0.3678}^*$ & 0.0073 & $\;\textbf{0.8113}^*$ & 0.52\% & $\;\textbf{0.4405}^*$ & 0.0041 \\
		
		\cline{1-11}
		\multicolumn{1}{c|}{\multirow{10}{*}{Convolution}} & \multicolumn{1}{c|}{\multirow{5}{*}{FiGNN}} & Scratch & 0.7923 & - & 0.3735 & - & 0.8094 & - & 0.4424 & - \\
		\multicolumn{1}{c|}{} & \multicolumn{1}{c|}{} & MF4UIP & 0.7925 & 0.03\% & 0.3728 & 0.0007 & 0.8117 & 0.28\% & 0.4405 & 0.0019 \\
		\multicolumn{1}{c|}{} & \multicolumn{1}{c|}{} & SCARF & 0.7941 & 0.23\% & 0.3717 & 0.0018 & $\underline{0.8118}$ & 0.30\% & $\underline{0.4404}$ & 0.0020 \\
		\multicolumn{1}{c|}{} & \multicolumn{1}{c|}{} & MFP (Ours) & $\;\underline{0.7971}^*$ & 0.61\% & $\;\underline{0.3697}^*$ & 0.0038 & 0.8117 & 0.28\% & $\underline{0.4404}$ & 0.0020 \\
		\multicolumn{1}{c|}{} & \multicolumn{1}{c|}{} & RFD (Ours) & $\;\textbf{0.7990}^*$ & 0.85\% & $\;\textbf{0.3684}^*$ & 0.0051 & $\;\textbf{0.8123}^*$ & 0.36\% & $\;\textbf{0.4395}^*$ & 0.0029 \\
		\cline{2-11}
		\multicolumn{1}{c|}{} & \multicolumn{1}{c|}{\multirow{5}{*}{FGCNN}} & Scratch & 0.7951 & - & 0.3727 & - & 0.8107 & - & 0.4413 & - \\
		\multicolumn{1}{c|}{} & \multicolumn{1}{c|}{} & MF4UIP & 0.7973 & 0.28\% & 0.3700 & 0.0027 & 0.8127 & 0.25\% & 0.4392 & 0.0021 \\
		\multicolumn{1}{c|}{} & \multicolumn{1}{c|}{} & SCARF & 0.7964 & 0.16\% & 0.3700 & 0.0027 & 0.8120 & 0.16\% & 0.4398 & 0.0015 \\
		\multicolumn{1}{c|}{} & \multicolumn{1}{c|}{} & MFP (Ours) & $\;\underline{0.7985}^*$ & 0.43\% & $\underline{0.3697}$ & 0.0030 & $\;\underline{0.8135}^*$ & 0.35\% & $\;\underline{0.4384}^*$ & 0.0029 \\
		\multicolumn{1}{c|}{} & \multicolumn{1}{c|}{} & RFD (Ours) & $\;\textbf{0.7992}^*$ & 0.52\% & $\;\textbf{0.3682}^*$ & 0.0045 & $\;\textbf{0.8139}^*$ & 0.39\% & $\;\textbf{0.4381}^*$ & 0.0032 \\
		
		\cline{1-11}
		\multicolumn{1}{c|}{\multirow{15}{*}{Product}} & \multicolumn{1}{c|}{\multirow{5}{*}{DeepFM}} & Scratch & 0.7924 & - & 0.3747 & - & 0.8103 & - & 0.4416 & - \\
		\multicolumn{1}{c|}{} & \multicolumn{1}{c|}{} & MF4UIP & 0.7970 & 0.58\% & 0.3692 & 0.0055 & 0.8109 & 0.07\% & 0.4414 & 0.0002 \\
		\multicolumn{1}{c|}{} & \multicolumn{1}{c|}{} & SCARF & 0.7992 & 0.86\% & 0.3684 & 0.0063 & 0.8117 & 0.17\% & 0.4400 & 0.0016 \\
		\multicolumn{1}{c|}{} & \multicolumn{1}{c|}{} & MFP (Ours) & $\;\underline{0.7998}^*$ & 0.93\% & $\underline{0.3680}$ & 0.0067 & $\;\underline{0.8126}^*$ & 0.28\% & $\;\underline{0.4392}^*$ & 0.0024 \\
		\multicolumn{1}{c|}{} & \multicolumn{1}{c|}{} & RFD (Ours) & $\;\textbf{0.8010}^*$ & 1.09\% & $\;\textbf{0.3671}^*$ & 0.0076 & $\;\textbf{0.8139}^*$ & 0.44\% & $\;\textbf{0.4380}^*$ & 0.0036 \\
		\cline{2-11}
		\multicolumn{1}{c|}{} & \multicolumn{1}{c|}{\multirow{5}{*}{xDeepFM}} & Scratch & 0.7967 & - & 0.3718 & - & 0.8112 & - & 0.4407 & - \\
		\multicolumn{1}{c|}{} & \multicolumn{1}{c|}{} & MF4UIP & 0.7982 & 0.19\% & 0.3691 & 0.0027 & 0.8130 & 0.22\% & 0.4390 & 0.0017 \\
		\multicolumn{1}{c|}{} & \multicolumn{1}{c|}{} & SCARF & $\underline{0.7992}$ & 0.31\% & $\underline{0.3685}$ & 0.0033 & 0.8126 & 0.17\% & 0.4395 & 0.0012 \\
		\multicolumn{1}{c|}{} & \multicolumn{1}{c|}{} & MFP (Ours) & 0.7989 & 0.28\% & $\underline{0.3685}$ & 0.0033 & $\;\underline{0.8145}^*$ & 0.41\% & $\;\underline{0.4374}^*$ & 0.0033 \\
		\multicolumn{1}{c|}{} & \multicolumn{1}{c|}{} & RFD (Ours) & $\;\textbf{0.8012}^*$ & 0.56\% & $\;\textbf{0.3671}^*$ & 0.0047 & $\;\textbf{0.8152}^*$ & 0.49\% & $\;\textbf{0.4367}^*$ & 0.0040 \\
		\cline{2-11}
		\multicolumn{1}{c|}{} & \multicolumn{1}{c|}{\multirow{5}{*}{DCNv2}} & Scratch & 0.7964 & - & 0.3727 & - & 0.8118 & - & 0.4403 & - \\
		\multicolumn{1}{c|}{} & \multicolumn{1}{c|}{} & MF4UIP & 0.7987 & 0.29\% & 0.3686 & 0.0041 & 0.8149 & 0.38\% & 0.4370 & 0.0033 \\
		\multicolumn{1}{c|}{} & \multicolumn{1}{c|}{} & SCARF & 0.8019 & 0.69\% & 0.3666 & 0.0061 & 0.8143 & 0.31\% & 0.4376 & 0.0027 \\
		\multicolumn{1}{c|}{} & \multicolumn{1}{c|}{} & MFP (Ours) & $\;\underline{0.8029}^*$ & 0.82\% & $\;\underline{0.3661}^*$ & 0.0066 & $\;\underline{0.8164}^*$ & 0.57\% & $\;\underline{0.4356}^*$ & 0.0047 \\
		\multicolumn{1}{c|}{} & \multicolumn{1}{c|}{} & RFD (Ours) & $\;\textbf{0.8037}^*$ & 0.92\% & $\;\textbf{0.3655}^*$ & 0.0072 & $\;\textbf{0.8165}^*$ & 0.58\% & $\;\textbf{0.4355}^*$ & 0.0048 \\
		\cline{1-11}
		\bottomrule
	\end{tabular}
	}
\end{table*}
\begin{figure*}[t]
    \centering
    \includegraphics[width=0.99\textwidth]{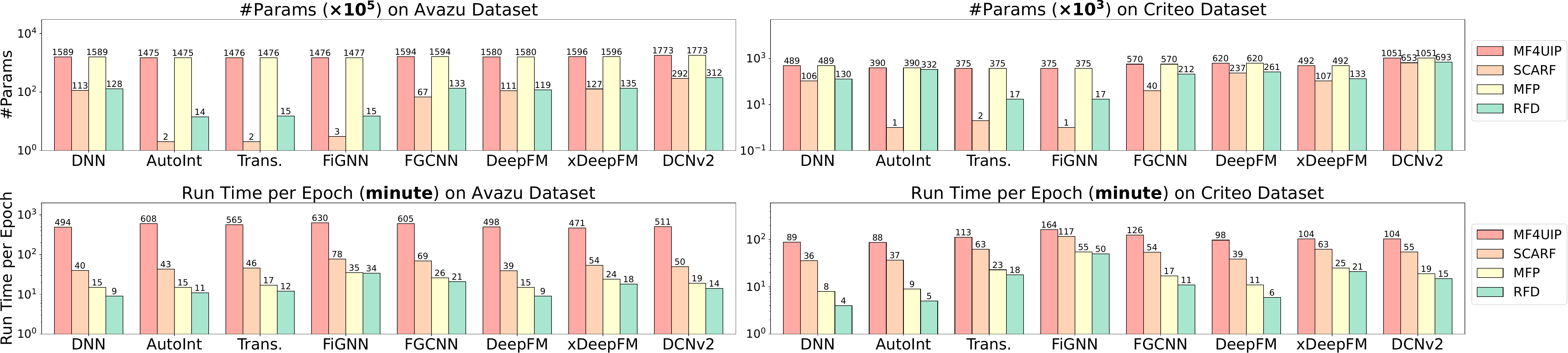}
    \caption{The \emph{model size} (Top) and \emph{run time per epoch} (Bottom) of different pretraining methods. 
    We perform logarithmic scale on the $y$ axis, and denote the original value on the top of each bar.
    The experiment is conducted on the same server with one GeForce RTX 3090 GPU.
    We only consider the learning parameters above the embedding layer for \emph{model size}. We consider the whole pretraining loop for \emph{run time per epoch}, including the corruption operations as well as the backpropagation.
    }
	\label{fig:efficiency}
\end{figure*}

\subsubsection{Base Models \& Baselines}
We evaluate the self-supervised pretraining methods on various base CTR models with three different feature interaction operators: (1) product operator, including DeepFM~\cite{DeepFM}, xDeepFM~\cite{xDeepFM}, and DCNv2~\cite{DCNv2}; (2) convolutional operator, including FiGNN~\cite{FiGNN} and FGCNN\cite{FGCNN}; (3) attention operator, including AutoInt~\cite{AutoInt} and Transformer~\cite{vaswani2017attention}. Additionally, we adopt the classical DNN model, which proves to be a strong base model for self-supervised learning in our experiments.
We compare our proposed MFP and RFD with two existing pretraining methods: MF4UIP~\cite{wang2020masked} and SCARF~\cite{bahri2021scarf}, which are chosen as representative generative and contrastive algorithms, respectively.

\subsubsection{Implementation Details}
We provide detailed implementation details in the supplementary material (\ie, Appendix~\ref{apx: implementation details}). 
The code is available\footnote{The PyTorch implementation is available at: \url{https://github.com/CHIANGEL/MAP-CODE}. The MindSpore implementation is available at: \url{https://gitee.com/mindspore/models/tree/master/research/recommend/MAP}}.

\subsection{Effectiveness Comparison (RQ1 \& RQ2)}
\label{sec:exp effectiveness}


We apply five training schemes on each base model, and report the results in Table~\ref{tab:overall performance}.
In the ``Scratch'' scheme, we train the randomly initialized base model from scratch (\ie, supervised learning).
In other schemes, we first pretrain the base model according to the corresponding method, and then finetune the pretrained model for CTR prediction (\ie, self-supervised learning).
From Table~\ref{tab:overall performance}, we can obtain the following observations:
\begin{itemize}[leftmargin=10pt]
    \item All the pretrain-finetune schemes can improve the performance of base model on both metrics by a large margin compared with the ``Scratch'' scheme, which demonstrates the effectiveness of self-supervised learning for CTR prediction.
    \item The product based CTR models (\eg, DCNv2), together with the DNN model, win the top places among the base models, when equipped with self-supervised learning.
    \item Among the pretrain-finetune schemes, MFP and RFD generally gain significant improvement over the two baseline pretraining methods except for few cases, and RFD can consistently achieve the best performance. 
\end{itemize}

In addition to the performance comparison above, we also find some interesting phenomena as follows:

\begin{itemize}[leftmargin=10pt]
    \item \emph{The pretrain-finetune scheme can greatly reduce the demand on model structure design for CTR prediction}. 
    To our surprise, although DNN model is inferior under the ``Scratch'' scheme, it gains huge improvement under the pretrain-finetune schemes (especially RFD) and wins the second place, outperforming a range of carefully designed CTR models.
    Such a phenomenon suggests that a simple MLP structure is capable of capturing useful feature crossing patterns from the multi-field categorical data with the help of self-supervised signals.
    \item \emph{The two training paradigms (supervised learning V.S. self-supervised learning) favor different types of model structures}.
    Previous works always seek better designs of model structures under the ``Scratch'' scheme. 
    However, we can observe several counterexamples that a good model under the ``Scratch'' scheme is relatively bad for pretrain-finetune schemes (\eg, FGCNN), or a bad model under the ``Scratch'' scheme achieves competitive performance for pretrain-finetune schemes (\eg, DNN). 
\end{itemize}

\noindent The above two phenomena bring about the research question of what model structures are more economic and effective for the pretrain-finetune schemes.
We give one conjecture about this topic here, and leave further studies as future works.
Our hypothesis is that the pretrain-finetune scheme might prefer the bit-wise feature interaction (\eg, DCNv2) to the field-wise feature interaction (\eg, Transformer). The bit-wise feature interaction enables larger model capacity to learn better feature crossing patterns during the pretraining stage. 

\subsection{Efficiency Analysis (RQ3)}
\label{sec:exp efficiency}
\begin{figure}[t]
    \centering
    \includegraphics[width=0.47\textwidth]{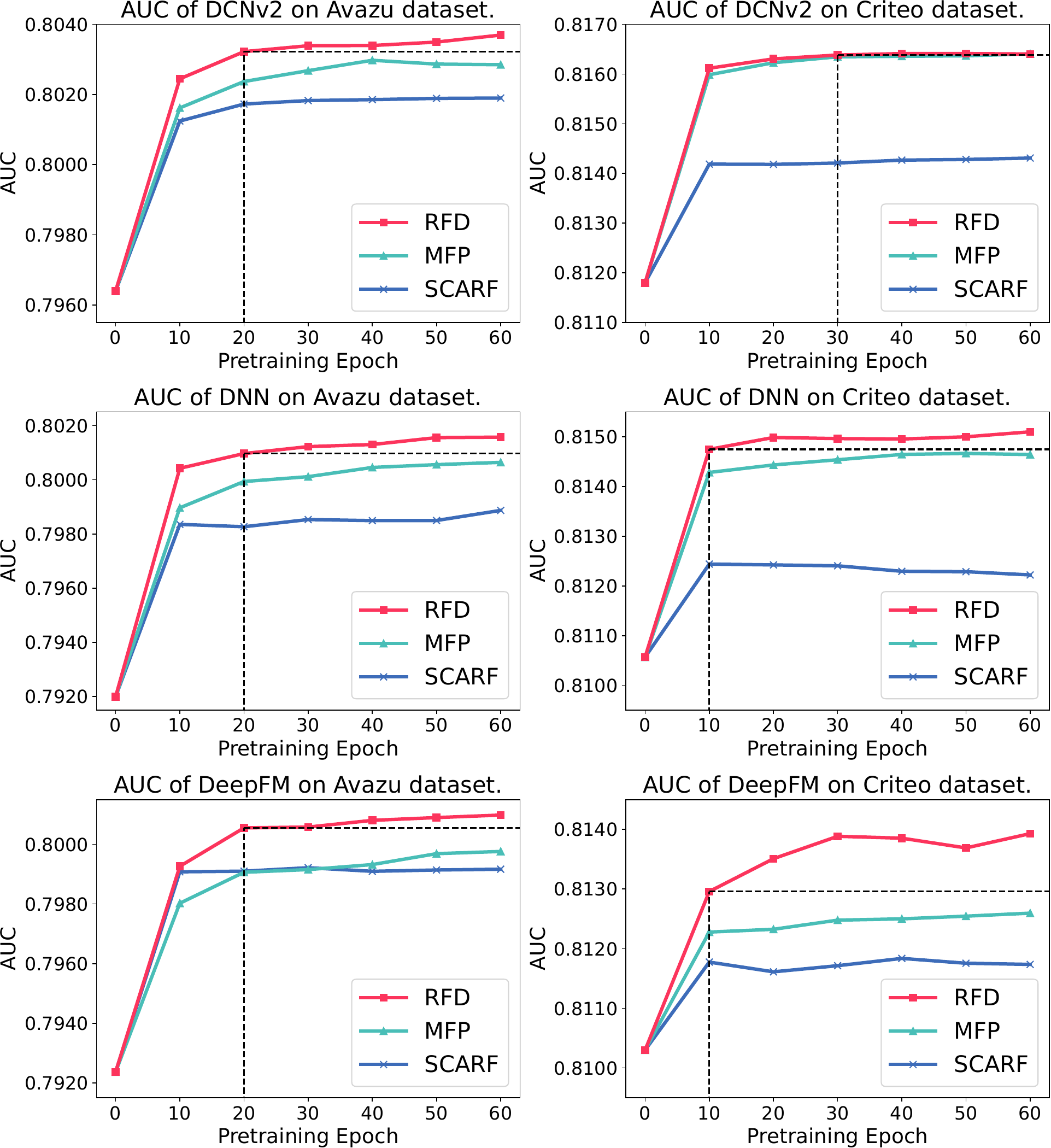}
    \caption{The AUC performance of three representative models (DCNv2, DNN, DeepFM) with different pretraining epochs on Avazu (left column) and Criteo (right column) datasets. 
    We use black dashed lines to illustrate how many epochs should RFD take to achieve a dominant performance over other methods.
    }
    \vspace{-10pt}
	\label{fig:RFD vs MFP vs SCARF}
\end{figure}

After validating the effectiveness of our proposed MFP and RFD, we conduct experiments to further analyze the efficiency of these pretraining methods from the following two perspectives:
\begin{itemize}[leftmargin=30pt]
    \item[\textbf{RQ3.1}] What is the complexity of each pretraining method?
    \item[\textbf{RQ3.2}] How many pretraining epochs should a method takes to achieve a certain performance (\ie, sample efficiency)?
\end{itemize}

For \textbf{RQ3.1}, we have already provided the complexity analysis in Section~\ref{sec:complexity analysis}. Following~\cite{wang2022enhancing}, we further empirically compare the \emph{model size} and \emph{run time per epoch} of different pretraining methods for different base CTR models in Figure~\ref{fig:efficiency}.
The experiments are conducted on the same server with one GeForce RTX 3090 GPU.
For fair comparison, we launch one pretraining at a time as a single process to exclusively possess all the computational resources.
We maintain the same structure of each base model for the four pretraining algorithms, and set the corrupt ratio $\gamma=0.3$.
Since different pretraining algorithms require different amounts of dynamic GPU memory, we choose a proper batch size from $\{256, 512, 1024, 2048, 4096\}$ to make full use of GPU memory.
From Figure~\ref{fig:efficiency}, we can obtain the following observations:

\begin{figure*}[t]
    \centering
    \includegraphics[width=0.99\textwidth]{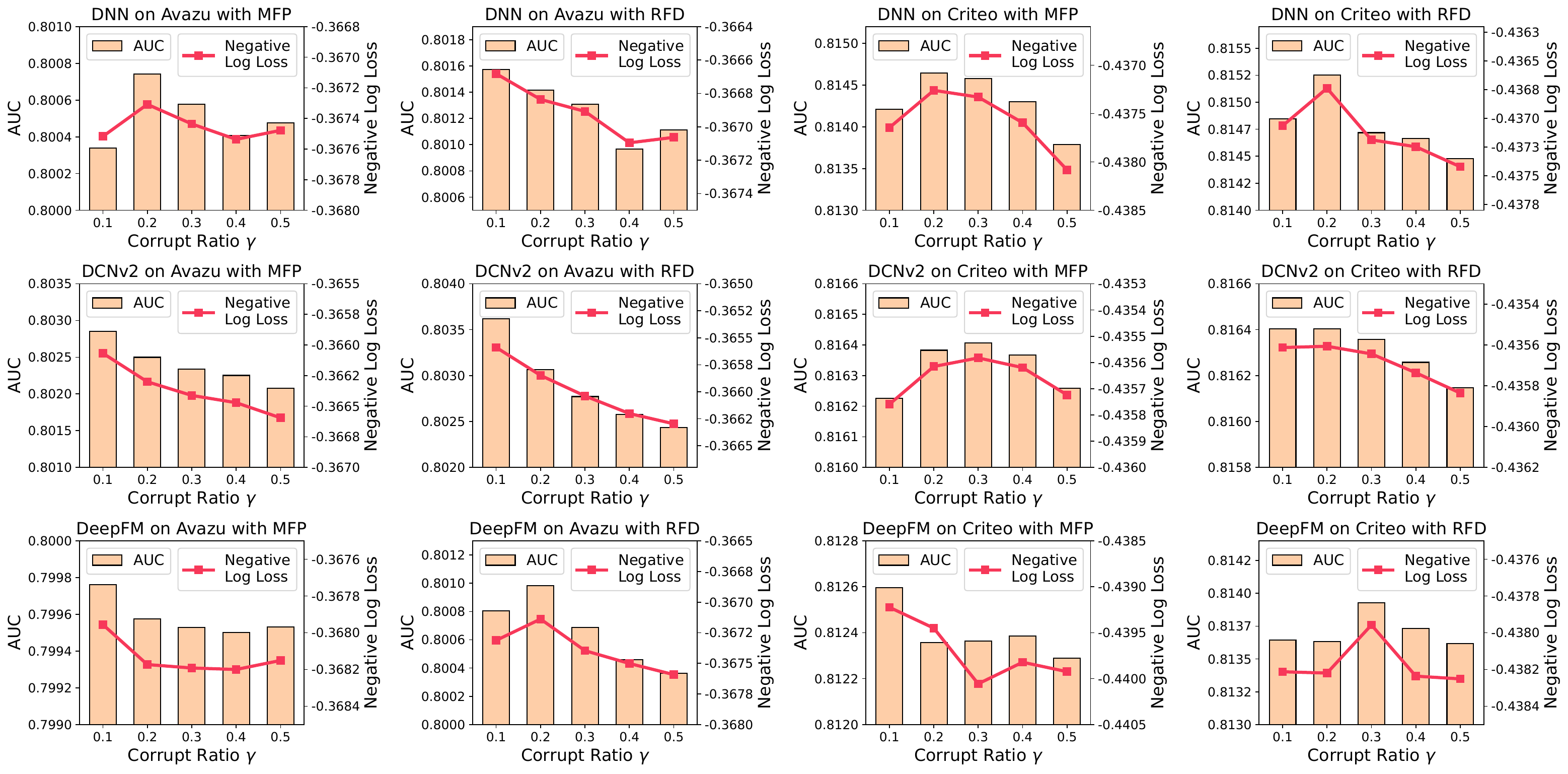}
    \caption{
    The hyperparameter study on corrupt ratio $\gamma$. 
    We give the AUC and negative log loss performance of DNN, DCNv2 and DeepFM with MFP and RFD methods on Avazu (left two columns) and Criteo (right two columns) datasets.
    }
	\label{fig:corrupt ratio}
\end{figure*}

\begin{itemize}[leftmargin=10pt]
    \item SCARF is relatively time-consuming, though it has minimal parameters. The reason is that SCARF, as a contrastive method, requires computing representations of different views from the same instance, which doubles the run time of training loops.
    \item Although MFP maintains similar amount of learning parameters compared with MF4UIP, it approximately exhibits an $18 \times$ speedup over MF4UIP in terms of run time per epoch since we adopt NCE to resolve the softmax overhead for million-level predictive spaces.
    \item RFD is the most efficient pretraining algorithm with the lowest complexity. It has relatively fewer learning parameters and the lowest run time per epoch, showing its simplicity and practicability for industrial applications.
\end{itemize}

Next, to study \textbf{RQ3.2}, we investigate the sample efficiency and give the AUC performance of base models under different pretraining epochs in Figure~\ref{fig:RFD vs MFP vs SCARF}. 
We choose DCNv2, DNN, and DeepFM as the representative base models due to the page limitation.
MF4UIP is excluded due to its tremendous cost of time per epoch.
Note that zero pretraining epoch indicates that we train the model from scratch without pretraining.
In Figure~\ref{fig:RFD vs MFP vs SCARF}, we can observe that MFP and RFD consistently achieve better performance over SCARF under different pretraining epochs.
Moreover, as illustrated by the black dashed lines, RFD can simply achieve the best performance with limited pretraining epochs (10$\sim$30), showing its superior sample efficiency for CTR prediction.

In summary, we validate the pretraining efficiency of our proposed methods (especially RFD), \ie, they can achieve better CTR performance with fewer learning parameters, higher throughput rates, and fewer pretraining epochs.

\subsection{Ablation \& Hyperparameter Study (RQ4)}

In this section, we analyze the impact of hyperparameters or components in MFP and RFD, including the corrupt ratio $\gamma$ for both MFP and RFD, the number of noise features $K$ in NCE for MFP, and the feature replacement strategy for RFD.
Similarly, we select DNN, DCNv2 and DeepFM as the representative base models due to the page limitation.

\subsubsection{Corrupt Ratio $\gamma$}

We select the value of corrupt ratio $\gamma$ from $\{0.1, 0.2, 0.3, 0.4, 0.5\}$, and show the impact in Figure~\ref{fig:corrupt ratio}. Both MFP and RFD favor a small corrupt ratio (\ie, 0.1$\sim$0.3). The reason is that the over-corruption caused by a large corrupt ratio may change the sample semantics and disturb the model pretraining.

\begin{table}[t]
	\centering
	\caption{
	The hyperparameter study on the number of noise features $K$ in NCE for MFP. 
    Different $K$s only result in 0.0003 performance fluctuation on each model.
	}
	\label{tab:ptNeg}
	\resizebox{0.48\textwidth}{!}{
	\renewcommand\arraystretch{1.15}
	\begin{tabular}{c|c|c|c c c c c}
		\toprule
		\cline{1-8} \multicolumn{1}{c|}{\multirow{2}{*}{Dataset}} & \multicolumn{1}{c|}{\multirow{2}{*}{Model}} & \multicolumn{1}{c|}{\multirow{2}{*}{Metric}} &   \multicolumn{5}{c}{the number of noise features $K$}  \\
		\cline{4-8}
		\multicolumn{1}{c|}{} & \multicolumn{1}{c|}{} & \multicolumn{1}{c|}{} & 10 & 25 & 50 & 75 & 100 \\
		
		\hline
		\multicolumn{1}{c|}{\multirow{6}{*}{Avazu}} & \multicolumn{1}{c|}{\multirow{2}{*}{DNN}} & AUC & 0.8006 & 0.8006 & 0.8005 & 0.8006 & 0.8007 \\
		\multicolumn{1}{c|}{} & \multicolumn{1}{c|}{} & Log Loss & 0.3674 & 0.3675 & 0.3674 & 0.3674 & 0.3673 \\
		\cline{2-8}
		\multicolumn{1}{c|}{} & \multicolumn{1}{c|}{\multirow{2}{*}{DCNv2}} & AUC & 0.8029 & 0.8029 & 0.8027 & 0.8027 & 0.8027 \\
		\multicolumn{1}{c|}{} & \multicolumn{1}{c|}{} & Log Loss & 0.3661 & 0.3661 & 0.3661 & 0.3662 & 0.3661 \\
		\cline{2-8}
		\multicolumn{1}{c|}{} & \multicolumn{1}{c|}{\multirow{2}{*}{DeepFM}} & AUC & 0.7998 & 0.7998 & 0.7999 & 0.7998 & 0.7999 \\
		\multicolumn{1}{c|}{} & \multicolumn{1}{c|}{} & Log Loss & 0.3680 & 0.3680 & 0.3678 & 0.3680 & 0.3680 \\
		
		\hline
		\multicolumn{1}{c|}{\multirow{6}{*}{Criteo}} & \multicolumn{1}{c|}{\multirow{2}{*}{DNN}} & AUC & 0.8145 & 0.8145 & 0.8146 & 0.8146 & 0.8147 \\
		\multicolumn{1}{c|}{} & \multicolumn{1}{c|}{} & Log Loss & 0.4374 & 0.4373 & 0.4372 & 0.4372 & 0.4372 \\
		\cline{2-8}
		\multicolumn{1}{c|}{} & \multicolumn{1}{c|}{\multirow{2}{*}{DCNv2}} & AUC & 0.8163 & 0.8164 & 0.8164 & 0.8165 & 0.8165 \\ \multicolumn{1}{c|}{} & \multicolumn{1}{c|}{\multirow{2}{*}{}} & Log Loss & 0.4357 & 0.4356 & 0.4356 & 0.4355 & 0.4355 \\
		\cline{2-8}
		\multicolumn{1}{c|}{} & \multicolumn{1}{c|}{\multirow{2}{*}{DeepFM}} & AUC & 0.8125 & 0.8126 & 0.8127 & 0.8126 & 0.8126 \\ \multicolumn{1}{c|}{} & \multicolumn{1}{c|}{\multirow{2}{*}{}} & Log Loss & 0.4392 & 0.4392 & 0.4390 & 0.4393 & 0.4392 \\
		\cline{1-8}
		\bottomrule
	\end{tabular}
	}
    \vspace{-10pt}
\end{table}

\begin{table}[t]
    \centering
    \caption{
    The feature replacement strategy variants for RFD pretraining method. It is worth noting that our proposed RFD adopts the field-frequency strategy.
    }
	\label{tab:feature replacement strategy variant}
    \includegraphics[width=0.48\textwidth]{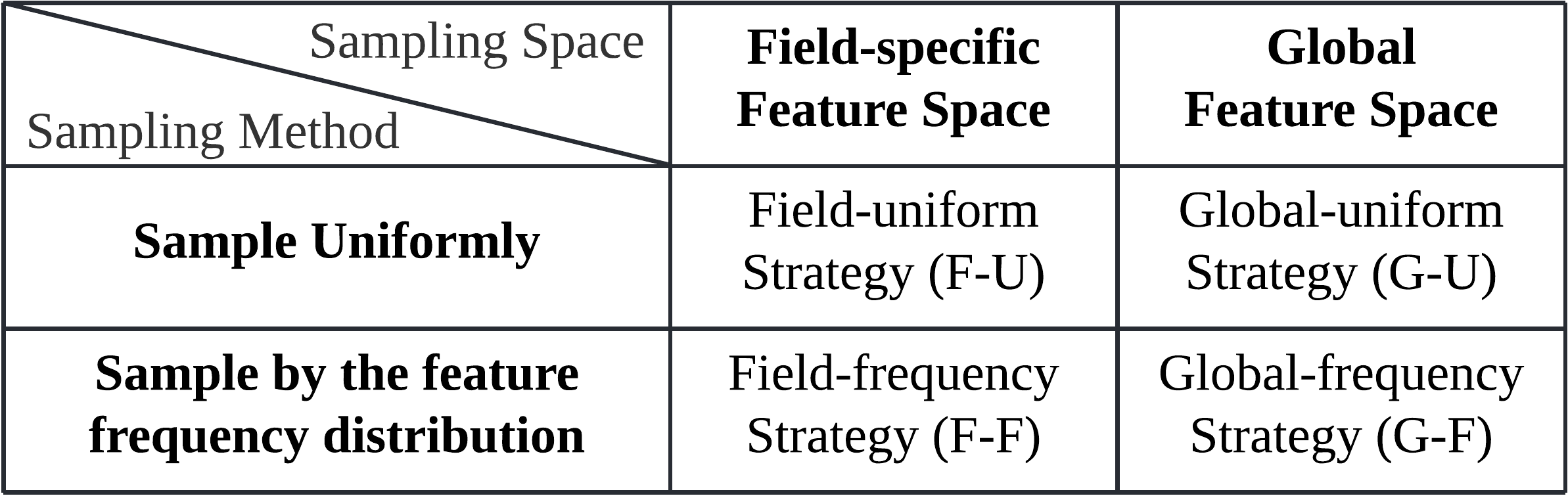}
    \vspace{-20pt}
\end{table}
\begin{table}[t]
	\centering
	\caption{
	The ablation study on feature replacement strategy for RFD. 
	F-F, F-U, G-F, G-U is short for field-frequency, field-uniform, global-frequency, global-uniform strategies, respectively.
	The best results are given in bold, while the second-best values are underlined.
	}
	\label{tab:feature replacement strategy}
	\resizebox{0.48\textwidth}{!}{
	\renewcommand\arraystretch{1.15}
	\begin{tabular}{c|c|c|c c c c}
		\toprule
		\cline{1-7} \multicolumn{1}{c|}{\multirow{2}{*}{Dataset}} & \multicolumn{1}{c|}{\multirow{2}{*}{Model}} & \multicolumn{1}{c|}{\multirow{2}{*}{Metric}} &   \multicolumn{4}{c}{Feature Replacement Strategy}  \\
		\cline{4-7}
		\multicolumn{1}{c|}{} & \multicolumn{1}{c|}{} & \multicolumn{1}{c|}{} & F-F (RFD) & F-U & G-F & G-U \\
		
		\hline
		\multicolumn{1}{c|}{\multirow{6}{*}{Avazu}} & \multicolumn{1}{c|}{\multirow{2}{*}{DNN}} & AUC & \textbf{0.8016} & \underline{0.8012} & 0.7996 & 0.7955 \\
		\multicolumn{1}{c|}{} & \multicolumn{1}{c|}{} & Log Loss & \textbf{0.3666} & \underline{0.3669} & 0.3680 & 0.3705 \\
		\cline{2-7}
		\multicolumn{1}{c|}{} & \multicolumn{1}{c|}{\multirow{2}{*}{DCNv2}} & AUC & \textbf{0.8037} & \underline{0.8035} & 0.8028 & 0.8016 \\
		\multicolumn{1}{c|}{} & \multicolumn{1}{c|}{} & Log Loss & \textbf{0.3655} & \underline{0.3656} & 0.3661 & 0.3669 \\
		\cline{2-7}
		\multicolumn{1}{c|}{} & \multicolumn{1}{c|}{\multirow{2}{*}{DeepFM}} & AUC & \textbf{0.8010} & \underline{0.8008} & 0.7991 & 0.7947 \\
		\multicolumn{1}{c|}{} & \multicolumn{1}{c|}{} & Log Loss & \textbf{0.3671} & \underline{0.3675} & 0.3688 & 0.3705 \\
		
		\hline
		\multicolumn{1}{c|}{\multirow{6}{*}{Criteo}} & \multicolumn{1}{c|}{\multirow{2}{*}{DNN}} & AUC & \textbf{0.8152} & \underline{0.8145} & 0.8118 & 0.8093 \\
		\multicolumn{1}{c|}{} & \multicolumn{1}{c|}{} & Log Loss & \textbf{0.4367} & \underline{0.4373} & 0.4402 & 0.4423 \\
		\cline{2-7}
		\multicolumn{1}{c|}{} & \multicolumn{1}{c|}{\multirow{2}{*}{DCNv2}} & AUC & \textbf{0.8165} & \underline{0.8163} & 0.8152 & 0.8131 \\ 
		\multicolumn{1}{c|}{} & \multicolumn{1}{c|}{} & Log Loss & \textbf{0.4355} & \underline{0.4357} & 0.4367 & 0.4387 \\
		\cline{2-7}
		\multicolumn{1}{c|}{} & \multicolumn{1}{c|}{\multirow{2}{*}{DeepFM}} & AUC & \textbf{0.8139} & \underline{0.8130} & 0.8100 & 0.8079 \\ 
		\multicolumn{1}{c|}{} & \multicolumn{1}{c|}{} & Log Loss & \textbf{0.4380} & \underline{0.4389} & 0.4417 & 0.4442 \\
		\cline{1-7}
		\bottomrule
	\end{tabular}
	}
    \vspace{-10pt}
\end{table}

\subsubsection{The Number of Noise Samples $K$}
We select the number of noise samples $K$ in NCE for MFP from $\{10, 25, 50, 75, 100\}$, and show the impact in Table~\ref{tab:ptNeg}. 
Surprisingly, the performance fluctuations of both metrics (\ie, AUC and log loss) brought by different $K$s are all within 0.0003, indicating that MFP is not sensitive to the number of noise samples $K$ in NCE.
A small number of noise features (10 noise features out of the million-level feature space) is sufficient for the model to learn effective feature crossing patterns and benefit the final CTR performance.

\subsubsection{Feature Replacement Strategy}
We investigate the impact of the feature replacement strategy in RFD, which needs to sample a replacer for the original feature.
We compare four different replacement strategy variants shown in Table~\ref{tab:feature replacement strategy variant}, and give the results in Table~\ref{tab:feature replacement strategy}. 
We observe that sampling by the feature frequency distribution is relatively better than uniform sampling. 
Moreover, sampling from the global feature space can greatly hurt the CTR performance, since an out-of-field replacer feature forms a simplistic pretext task where the model can easily detect the replaced field.
The model might overfit the easy pretext task during the pretraining, and thus lose the generalization ability for downstream CTR prediction task.

\section{Related Work}

\subsection{Click-through Rate Prediction}

The click-through rate (CTR) prediction serves as a core function module in various personalized online services, including online advertising, recommender systems, and web search, etc~\cite{zhang2021deep,lin2023can}. 
With the rise of deep learning, many deep neural CTR models have been recently proposed. The core idea of them is to capture the feature interactions, which indicates the combination relationships of multiple features.
The deep CTR models usually leverage both implicit and explicit feature interactions. While the implicit feature interactions are captured by a deep neural network (DNN), the explicit feature interactions are modeled by a specially designed learning function.
According to the explicit feature interaction operators, these deep CTR models can be mainly classified into three categories: (1) product operator, (2) convolutional operator, and (3) attention operator.

\textbf{Product Operator}.
The product-based CTR models originate from classical shallow models such as FM~\cite{FM} and POLY2~\cite{POLY2}.
FFM~\cite{FFM} and NFM~\cite{NFM} are variants of FM, where the second-order feature interactions are captured by the inner product of feature embeddings.
DeepFM~\cite{DeepFM} and PNN~\cite{PNN} combine the FM layer with DNNs for higher-order feature interactions.
PIN~\cite{PIN} further extends PNN by introducing a network-in-network structure to replace the inner product interaction function. 
FiBiNET~\cite{FiBiNET} introduces the SENET mechanism~\cite{hu2018squeeze} to learn the weights of
features dynamically before product interactions.
Moreover, DCN~\cite{DCNv1}, xDeepFM~\cite{xDeepFM}, DCNv2~\cite{DCNv2} are proposed for the explicit high-order feature interaction modeling by applying product-based feature interactions at each layer explicitly. 
Therefore, the order of feature interactions to be modeled increases at each layer and is determined by the layer depth.

\textbf{Convolutional Operator}.
Apart from the product operator, Convolutional Neural Networks (CNN) and Graph Convolutional Networks (GCN) are also explored for feature interaction modeling in CTR prediction.
CCPM~\cite{CCPM} is the first work adopts the CNN module for CTR prediction. 
However, CCPM can only learn part of feature interactions between adjacent features since it is sensitive to the field order. 
FGCNN~\cite{FGCNN} improves CCPM by introducing a recombination layer to model non-adjacent features. 
FiGNN~\cite{FiGNN} treats the multi-field categorical data as a fully connected graph, where each field serves as a graph node and feature interactions are captured via graph propagation.

\textbf{Attention Operator}.
AFM~\cite{AFM} improves FM by leveraging an additional attention network to allow feature interactions to contribute differently to the final CTR prediction.
AutoInt~\cite{AutoInt} utilizes a multi-head self-attentive neural network with residual connections to explicitly model the feature interactions with different orders.
InterHAt~\cite{InterHAt} combines a transformer network with multiple attentional aggregation layers for feature interaction learning.
These attention-based CTR models can also provide explainable prediction via attention weights.


\subsection{Self-supervised Learning}

Self-supervised learning has achieved great success in Nature Language Processing (NLP)~\cite{devlin2018bert,clark2020electra} and Computer Vision (CV)~\cite{MoCo}. 
They usually adopt a pretrain-finetune scheme, where we first pretrain an encoder based on pretext tasks, and then finetune a model initialized by the pretrained encoder for downstream tasks.
According to the pretext task, self-supervised learning can be mainly classified into two categories: (1) contrastive methods and (2) generative methods.
\emph{Contrastive methods} learn a latent space to draw positive samples together (\eg, different views of the same image) and push apart negative samples (\eg, images from different categories)~\cite{dwibedi2021little}. Numerous techniques are proposed to promote the performance of contrastive methods such as data augmentation~\cite{zhang2017mixup,hao2022mixgen}, contrastive losses~\cite{chen2020simple,chen2020big}, momentum encoders~\cite{MoCo,chen2020improved,grill2020bootstrap}, and memory banks~\cite{tian2020contrastive,wu2018unsupervised}. 
\emph{Generative methods}~\cite{devlin2018bert,he2022masked,brown2020language,yang2019xlnet} reconstruct the original input sample from the corrupted one.
For example, BERT~\cite{devlin2018bert} requires the model to recover the masked token from the corrupted sentences.
Besides, ELECTRA~\cite{clark2020electra} adopts an adversarial structure and pretrains the model as a discriminator to predict corrupted tokens, which is proven to be more sample-efficient.

In recommender systems, many works apply self-supervised learning to user behavior sequences~\cite{guo2022miss,sun2019bert4rec,liu2021contrastive,xie2020contrastive,wang2022contrastvae}, manually designed graph data~\cite{wang2021pre,hao2021pre,meng2021graph}, or multi-modal data~\cite{liu2022multi,liu2021pre}.
They explore the pretraining methods for better representations to further enhance the recommendation performance with enriched side information.
However, these sequential/graph/multi-modal methods are essentially incompatible with the CTR data, \ie, multi-field categorical data format.

In CTR prediction, there exist works~\cite{wang2022cl4ctr,pan2021click,yao2021self} that incorporate the self-supervised signals in a semi-supervised manner, where the cross-entropy loss is jointly optimized with an auxiliary loss in one stage.
As for CTR pretraining methods, VIME~\cite{yoon2020vime} proposes a semi-supervised learning algorithm to learn a predictive function based on the frozen pretrained encoder.
MF4UIP~\cite{wang2020masked} leverages the BERT framework~\cite{devlin2018bert} for user intent prediction.
SCARF~\cite{bahri2021scarf} adopts SimCLR framework~\cite{chen2020simple} and InfoNCE loss~\cite{oord2018representation} to pretrain the model in a contrastive manner.
Compared with these works, our proposed MFP and RFD methods are more scalable for industrial applications and achieve the state-of-art performance in terms of both effectiveness and efficiency for CTR prediction.


\section{Conclusion}

In this paper, we propose a Model-agnostic Pretraining (MAP) framework that applies feature corruption and recovery on multi-field categorical data for CTR prediction. 
Based on different strategies of corruption and recovery, we derive two practical algorithms: masked feature prediction (MFP), and replaced feature detection (RFD).
Extensive experiments show that MFP and RFD achieve new state-of-the-art performance in terms of both effectiveness and efficiency for CTR prediction.
For future work, a promising direction is to explore what model structures are more suitable for self-supervised paradigm, since we find different models receive quite different performance gains combined with self-supervised learning in Section~\ref{sec:exp effectiveness}.
Furthermore, we will investigate on the possible saturation of downstream CTR performance as the pretraining volume grows (\eg, from million to billion or even more).

\begin{acks}
The SJTU team is supported by Shanghai Municipal Science and Technology Major Project (2021SHZDZX0102) and National Natural Science Foundation of China (62177033). The work is also sponsored by Huawei Innovation Research Program. We thank MindSpore~\cite{mindspore} for the partial support of this work.
\end{acks}

\bibliographystyle{ACM-Reference-Format}
\bibliography{acmart}

\appendix



\section{Implementation Details}
\label{apx: implementation details}

In this section, we describe the implementation details for our empirical experiments.
We conduct both supervised learning (\ie, the ``Scratch'' scheme) and self-supervised learning (\ie, four pretrain-finetune schemes) over different base models (\ie, CTR model).
We first introduce the model configuration for each base model, and then give the training settings for supervised learning and self-supervised learning, respectively.
Finally, we describe how to employ the pretraining methods for the assembled models (\eg, DCNv2 and DeepFM).

\subsection{Configuration for Base Models}
We choose the embedding size from $\{16, 32, 64\}$. 
The dropout rate is selected from $\{0.0, 0.1, 0.2\}$.
We utilize one linear layer after the feature interaction layer to make the final CTR prediction.
Unless stated otherwise, we adopt ReLU as the activation function.
The model-specific hyperparameter settings for base models are as follows:
\begin{itemize}[leftmargin=10pt]
    \item \textbf{DNN}. We select the size of DNN layer from $\{1000, 2000\}$, and the number of DNN layers from $\{3, 6, 9, 12\}$.
    \item \textbf{AutoInt}. We select the number of attention layers from $\{3, 6, 9, 12\}$. The number of attention heads per layer and the attention size are set to $1$ and $64$, respectively.
    \item \textbf{Transformer}. We select the number of layers from $\{3, 6, 9, 12\}$. The number of attention heads is set to $1$, and the intermediate size of feed-forward network is set to quadruple the embedding size. We also try both the post-norm~\cite{vaswani2017attention} and pre-norm~\cite{xiong2020layer} structure.
    \item \textbf{FiGNN}. We select the number of layers from $\{3, 6, 9, 12\}$, and apply residual connection for the graph layers.
    \item \textbf{FGCNN}. We maintain 4 tanh-activated convolutional layers with a kernel size of 7 and pooling size of 2 for each layer. The number of channels for each layer is set to $8, 10, 12, 14$, respectively. The numbers of channels for recombination layers are all set to $3$.
    \item \textbf{DeepFM}. We select the size of DNN layer from $\{1000, 2000\}$, and the number of DNN layers from $\{3, 6, 9, 12\}$.
    \item \textbf{xDeepFM}. We choose the number of CIN layers from $\{2, 3, 4, 5\}$, and the number of units per CIN layer is set to $25$. We select the size of DNN layer from $\{1000, 2000\}$, and the number of DNN layers from $\{3, 6, 9, 12\}$.
    \item \textbf{DCNv2}. We select the size of DNN layer from $\{1000, 2000\}$, and the number of DNN layers from $\{3, 6, 9, 12\}$.
    We force the CrossNet module to have the same number of layer as the DNN network.
\end{itemize}

\subsection{Settings for Supervised Learning}
We train each base model from scratch based on click signals without pretraining.
We adopt the Adam optimizer with weight decay rate selected from $\{0.01, 0.05\}$.
The batch size is $4096$, and the learning rate is chosen from $\{10^{-3}, 7\times10^{-4}, 5\times10^{-4}\}$ without decay.
We adopt early stop if the AUC performance on the validation set stops increasing for two consecutive epochs.
Finally, we choose the model at the iteration with the highest validation AUC performance for evaluation in the test set.

\subsection{Settings for Self-supervised Learning}
For self-supervised learning paradigm, we implement four different pretraining methods for CTR prediction. We give the settings for the pretraining and finetuning stages as follows.

\subsubsection{Pretraining Stage}
We adopt the Adam optimizer with the weight decay rate of $0.05$.
The learning rate is initialized at $10^{-3}$, and is scheduled by cosine decay. 
There is no warm-up epochs during the pretraining.
The corrupt ratio is selected from $\{0.1, 0.2, 0.3\}$.
We adopt a two-layer MLP with 32 hidden units for the (field-wise) output layer.
The method-specific settings are as follows:
\begin{itemize}[leftmargin=10pt]
    \item \textbf{MF4UIP}. We set the batch size to $256$ and only pretrain each base model for $5$ epoch, due to its tremendous cost of GPU memory and throughput time.
    \item \textbf{SCARF}. We set the batch size to $2048$, and pretrain each base model for $60$ epochs. The temperature in InfoNCE loss is $1.0$.
    \item \textbf{MFP}. We set the batch size to $4096$, and pretrain each base model for $60$ epochs. The number of noise samples $K$ for NCE is $25$.
    \item \textbf{RFD}.  We set the batch size to $4096$, and pretrain each base model for $60$ epochs.
\end{itemize}

\subsubsection{Finetuning Stage}
The batch size is set to $4096$.
The initial learning rate is selected from $\{10^{-3}, 7\times10^{-4}, 5\times10^{-4}\}$, and is scheduled by cosine decay.
The total finetuning epoch is chosen from $\{1, 2, 3, 4\}$.
We adopt the Adam optimizer and choose the weight decay rate from $\{0.01, 0.05\}$.
We choose the model at the iteration with the highest validation AUC performance for evaluation in the test set.

\subsection{How to Pretrain the Assembled Models?}

The deep CTR models usually leverage the parallel structure to incorporate the explicit feature interaction model with the DNN module as follows:
\begin{equation}
    p(y_i=1|x_i)=f(x_i)+\operatorname{MLP}(x_i),
\end{equation}
where $f(\cdot)$ is the explicit feature interaction model.
For examples, DCNv2 assembles DNN and CrossNet. DeepFM assembles DNN and FM.
We abstract them as \emph{assembled models} that add up outputs from multiple ($\ge 2$) modules for final CTR prediction.

Suppose we have $N$ modules to be assembled. 
Each of them produces one vector $q_{i,k}\,(k=1,\dots,N)$ for the input sample $x_i$ before the last prediction layer.
If the module is a shallow network that outputs a scalar (\eg, LR, FM), we simply denote it as a vector that has only one dimension.
Then, we get the compact representation $q_i$ from the assembled model by concatenating the $N$ output vectors:
\begin{equation}
    q_i=[q_{i,1} \oplus q_{i,2} \oplus \cdots \oplus q_{i,N}],
\end{equation}
where $\oplus$ is the vector concatenation.
After obtaining the compact representation $q_i$, we can follow the methodology described in Section~\ref{sec:method} to pretrain and finetune the assembled model.

\begin{table*}[t]
    \centering
    \caption{
        The ablation study on the finetuning strategies (\ie. whether freeze the embedding layer and feature interaction layer or not). The best results are given in bold, while the second-best values are underlined.
    }
    \label{tab:finetune strategy}
    \renewcommand\arraystretch{1.15}
    \begin{tabular}{c|c|c c|c c|c c}
        \toprule
        \cline{1-8} 
        \multicolumn{1}{c|}{\multirow{2}{*}{Pretraining Strategy}} & \multicolumn{1}{c|}{\multirow{2}{*}{Base Model}} & \multicolumn{2}{c|}{Finetuning Strategy} &   \multicolumn{2}{c}{Avazu} &  \multicolumn{2}{c}{Criteo} \\
        \cline{3-8}
        \multicolumn{1}{c|}{} & \multicolumn{1}{c|}{} & Update Embed & Update FI & AUC & Log Loss & AUC & Log Loss \\
        
        \hline
        \multicolumn{1}{c|}{\multirow{12}{*}{MFP}} & \multicolumn{1}{c|}{\multirow{4}{*}{DNN}} & \ding{52} & \ding{52} & \textbf{0.8006} & \textbf{0.3675} & \textbf{0.8145} & \textbf{0.4373} \\
        \multicolumn{1}{c|}{} & \multicolumn{1}{c|}{} & \ding{52} & \ding{56} & 0.7958 & 0.3706 & 0.8083 & 0.4433 \\
        \multicolumn{1}{c|}{} & \multicolumn{1}{c|}{} & \ding{56} & \ding{52} & 0.7749 & 0.3824 & 0.8059 & 0.4451 \\
        \multicolumn{1}{c|}{} & \multicolumn{1}{c|}{} & \ding{56} & \ding{56} & 0.7325 & 0.4052 & 0.7611 & 0.4812 \\
        \cline{2-8}
        \multicolumn{1}{c|}{} & \multicolumn{1}{c|}{\multirow{4}{*}{DCNv2}} & \ding{52} & \ding{52} & \textbf{0.8029} & \textbf{0.3661} & \textbf{0.8164} & \textbf{0.4356} \\
        \multicolumn{1}{c|}{} & \multicolumn{1}{c|}{} & \ding{52} & \ding{56} & 0.8001 & 0.3679 & 0.8109 & 0.4409 \\
        \multicolumn{1}{c|}{} & \multicolumn{1}{c|}{} & \ding{56} & \ding{52} & 0.7767 & 0.3815 & 0.8103 & 0.4412 \\
        \multicolumn{1}{c|}{} & \multicolumn{1}{c|}{} & \ding{56} & \ding{56} & 0.7396 & 0.4019 & 0.7710 & 0.4739 \\
        \cline{2-8}
        \multicolumn{1}{c|}{} & \multicolumn{1}{c|}{\multirow{4}{*}{DeepFM}} & \ding{52} & \ding{52} & \textbf{0.7998} & \textbf{0.3680} & \textbf{0.8126} & \textbf{0.4392} \\
        \multicolumn{1}{c|}{} & \multicolumn{1}{c|}{} & \ding{52} & \ding{56} & 0.7951 & 0.3710 & 0.8057 & 0.4458 \\
        \multicolumn{1}{c|}{} & \multicolumn{1}{c|}{} & \ding{56} & \ding{52} & 0.7778 & 0.3808 & 0.8050 & 0.4461 \\
        \multicolumn{1}{c|}{} & \multicolumn{1}{c|}{} & \ding{56} & \ding{56} & 0.7580 & 0.3924 & 0.7762 & 0.4717 \\
        
        \hline
        \multicolumn{1}{c|}{\multirow{12}{*}{RFD}} & \multicolumn{1}{c|}{\multirow{4}{*}{DNN}} & \ding{52} & \ding{52} & \textbf{0.8016} & \textbf{0.3666} & \textbf{0.8152} & \textbf{0.4367} \\
        \multicolumn{1}{c|}{} & \multicolumn{1}{c|}{} & \ding{52} & \ding{56} & 0.7981 & 0.3691 & 0.8083 & 0.4434 \\
        \multicolumn{1}{c|}{} & \multicolumn{1}{c|}{} & \ding{56} & \ding{52} & 0.7763 & 0.3816 & 0.8065 & 0.4445 \\
        \multicolumn{1}{c|}{} & \multicolumn{1}{c|}{} & \ding{56} & \ding{56} & 0.6775 & 0.4269 & 0.7611 & 0.4812 \\
        \cline{2-8}
        \multicolumn{1}{c|}{} & \multicolumn{1}{c|}{\multirow{4}{*}{DCNv2}} & \ding{52} & \ding{52} & \textbf{0.8037} & \textbf{0.3655} & \textbf{0.8165} & \textbf{0.4355} \\
        \multicolumn{1}{c|}{} & \multicolumn{1}{c|}{} & \ding{52} & \ding{56} & 0.8011 & 0.3675 & 0.8112 & 0.4406 \\
        \multicolumn{1}{c|}{} & \multicolumn{1}{c|}{} & \ding{56} & \ding{52} & 0.7776 & 0.3811 & 0.8098 & 0.4416 \\
        \multicolumn{1}{c|}{} & \multicolumn{1}{c|}{} & \ding{56} & \ding{56} & 0.7355 & 0.4041 & 0.7639 & 0.4791 \\
        \cline{2-8}
        \multicolumn{1}{c|}{} & \multicolumn{1}{c|}{\multirow{4}{*}{DeepFM}} & \ding{52} & \ding{52} & \textbf{0.8010} & \textbf{0.3671} & \textbf{0.8139} & \textbf{0.4380} \\
        \multicolumn{1}{c|}{} & \multicolumn{1}{c|}{} & \ding{52} & \ding{56} & 0.7961 & 0.3705 & 0.8066 & 0.4450 \\
        \multicolumn{1}{c|}{} & \multicolumn{1}{c|}{} & \ding{56} & \ding{52} & 	0.7804 & 0.3794 & 0.8038 & 0.4472 \\
        \multicolumn{1}{c|}{} & \multicolumn{1}{c|}{} & \ding{56} & \ding{56} & 0.7577 & 0.3931 & 0.7861 & 0.4627\\
        
        \cline{1-8}
        \bottomrule
    \end{tabular}
\end{table*}

\section{Additional Experiments}

\subsection{Finetuning Strategy}

Since the embedding layer and feature interaction (FI) layer will be further updated during the finetuning stage. We provide an additional ablation study to investigate the influence of different finetuning strategies (\ie, freezing different parts of CTR models during finetuning). We choose DCNv2, DNN, and DeepFM as the representative models, and study the effect for both MFP and RFD tasks. The results are reported in Table~\ref{tab:finetune strategy}.

From Table~\ref{tab:finetune strategy}, we observe that either freezing the embedding layer or the feature interaction layer will badly hurt the final performance for all base models and pretraining methods. This indicates that there exists gap between the pretraining objective and CTR prediction task. The pretrained parameters provide a useful warm-up initialization, but still require further updating for the downstream CTR prediction task.

\subsection{Joint Pretraining of MFP and RFD}

We conduct ablation experiments to further analyze the effect of joint pretraining of our proposed MFP and RFD, where the loss function for each input $x_i$ is:
\begin{equation}
    L_i^{Joint}=\alpha \times L_i^{MFP} + (1 - \alpha) \times L_i^{RFD},
\end{equation}
where $\alpha$ is a hyperparameter to balance the loss terms from MFP and RFD.
We set other hyperparameters to be the same as the best configuration in the RFD pretrain-finetune scheme for simplicity, and apply grid search to select $\alpha$ from $\{0.1, 0.3, 0.5, 0.7, 0.9\}$. DCNv2, DNN, DeepFM are chosen as the representative base models. The results are reported in Table~\ref{tab:joint MFP RFD}.

\begin{table}[h]
    \centering
    \caption{
        The ablation study on the joint pretraining of MFP and RFD.
    The best results are given in bold, while the second-best values are underlined.
    }
    \label{tab:joint MFP RFD}
    \renewcommand\arraystretch{1.1}
    \begin{tabular}{c|c|c|c c c}
        \toprule
        \cline{1-6} \multicolumn{1}{c|}{\multirow{2}{*}{Dataset}} & \multicolumn{1}{c|}{\multirow{2}{*}{Model}} & \multicolumn{1}{c|}{\multirow{2}{*}{Metric}} &   \multicolumn{3}{c}{Pretraining Strategy}  \\
        \cline{4-6}
        \multicolumn{1}{c|}{} & \multicolumn{1}{c|}{} & \multicolumn{1}{c|}{} & MFP & RFD & Joint \\
        
        \hline
        \multicolumn{1}{c|}{\multirow{6}{*}{Avazu}} & \multicolumn{1}{c|}{\multirow{2}{*}{DNN}} & AUC & 0.8006 & \underline{0.8016} & \textbf{0.8017} \\
        \multicolumn{1}{c|}{} & \multicolumn{1}{c|}{} & Log Loss & 0.3675 & \textbf{0.3666} & \underline{0.3667} \\
        \cline{2-6}
        \multicolumn{1}{c|}{} & \multicolumn{1}{c|}{\multirow{2}{*}{DCNv2}} & AUC & 0.8029 & \textbf{0.8037} & \underline{0.8035} \\
        \multicolumn{1}{c|}{} & \multicolumn{1}{c|}{} & Log Loss & 0.3661 & \textbf{0.3655} & \underline{0.3660} \\
        \cline{2-6}
        \multicolumn{1}{c|}{} & \multicolumn{1}{c|}{\multirow{2}{*}{DeepFM}} & AUC & 0.7998 & \underline{0.8010} & \textbf{0.8015} \\
        \multicolumn{1}{c|}{} & \multicolumn{1}{c|}{} & Log Loss & 0.3680 & \underline{0.3671} & \textbf{0.3668} \\
        
        \hline
        \multicolumn{1}{c|}{\multirow{6}{*}{Criteo}} & \multicolumn{1}{c|}{\multirow{2}{*}{DNN}} & AUC & 0.8145 & \underline{0.8152} & \textbf{0.8153} \\
        \multicolumn{1}{c|}{} & \multicolumn{1}{c|}{} & Log Loss & 0.4373 & \underline{0.4367} & \textbf{0.4366} \\
        \cline{2-6}
        \multicolumn{1}{c|}{} & \multicolumn{1}{c|}{\multirow{2}{*}{DCNv2}} & AUC & \underline{0.8164} & \textbf{0.8165} & \underline{0.8164} \\ 
        \multicolumn{1}{c|}{} & \multicolumn{1}{c|}{} & Log Loss & \underline{0.4356} & \textbf{0.4355} & \underline{0.4356} \\
        \cline{2-6}
        \multicolumn{1}{c|}{} & \multicolumn{1}{c|}{\multirow{2}{*}{DeepFM}} & AUC & 0.8126 & \underline{0.8139} & \textbf{0.8145} \\ 
        \multicolumn{1}{c|}{} & \multicolumn{1}{c|}{} & Log Loss & 0.4392 & \underline{0.4380} & \textbf{0.4374} \\
        \cline{1-6}
        \bottomrule
    \end{tabular}
\end{table}

Our observation and discussion towards the results in Table~\ref{tab:joint MFP RFD} are in three folds:
\begin{itemize}[leftmargin=10pt]
    \item The joint method could achieve slightly better (or comparable) performance compared to the best pretraining method RFD.
    \item It is worth noting that the joint pretraining method consistently reaches the best performance with $\alpha = 0.1$. This indicates that RFD surpasses MFP and mainly contributes to the performance improvement.
    \item In addition, since the corruption and recovery strategies for MFP and RFD are different, the joint training method requires two forward/backward propagations for each data instance, which greatly increases the training cost (\eg, GPU memory usage and run time per epoch).
\end{itemize}

Although the joint training method might achieve better performance with finer-grained hyperparameter search, we think RFD is still a more elegant and practical pretraining method in terms of both effectiveness and efficiency.

\end{document}